\documentclass[aps,pra, amsmath,amssymb,amsfonts,floatfix,reprint]{revtex4-1}
\usepackage[utf8]{inputenc}
\usepackage[pdftex]{graphicx}
\usepackage{siunitx}
\usepackage{placeins}
\begin{document}
\title{Inverse Spin Hall Effect in $\text{Ni}_{81} \text{Fe}_{19}$ / Normal Metal Bilayers}
\author{M. Obstbaum, M. Härtinger, T. Meier, F. Swientek, C. H. Back and G. Woltersdorf}
\affiliation{Institut für Experimentelle und Angewandte Physik, Universität Regensburg, Germany}
\begin{abstract}
Spin pumping in ferromagnets provides a source of pure spin currents. Via the inverse spin Hall effect  a spin current is converted into a charge current and a corresponding detectable DC-voltage. The ratio of injected spin current to resulting charge current is given by the spin Hall angle. However, the number of experiments more or less equals the number of different values for spin Hall angles, even for the most studied normal metal platinum. This publication provides a full study of inverse spin Hall effect  and anisotropic magnetoresistance  for different $\text{Ni}_\text{81} \text{Fe}_\text{19}$(Py) / normal metal bilayers using a coplanar waveguide structure. Angle and frequency dependent measurements strongly suggest that spin pumping and inverse spin Hall effect  can be used to quantify spin Hall angles only if certain conditions are met. Ruling out the anisotropic magnetoresistance as a parasitic voltage generating effect measurements of the inverse spin Hall effect  in Py/Pt and Py/Au yield spin Hall angles of $0.09$ and $0.008$ respectively. Furthermore, DC-voltages at ferromagnetic resonance for Py/Pt are studied as a function of temperature and the results are compared to theoretical models.

\end{abstract}
\maketitle
In 2002 Tserkovnyak, Brataas and Bauer \cite{Tserkovnyak2002PRL,Tserkovnyak2002PRB} proposed the spin pumping mechanism to inject a pure spin current from a ferromagnet into an adjacent normal metal which can act as a spin sink. This effect can be evidenced in linewidth broadening of ferromagnetic resonance (FMR) curves \cite{Mizukami2001,Urban2001PRL,Heinrich2003PRL}. The inverse spin Hall effect (ISHE) offers the possiblity to detect pure spin currents generated by spin pumping electrically \cite{Saitoh2006}. In materials with reasonably large spin orbit coupling ISHE converts a pure spin current into a detectable charge current \cite{Dyakonov1971,Hirsch1999,Zhang2000}. The efficiency of this conversion process can be quantified by the spin Hall angle $\alpha_\text{SH}$.
 A number of materials have been investigated with respect to the spin Hall angle in mostly two experimental configurations: non local lateral spin valves (NLSV) and dynamic spin injection related to ferromagnetic resonance (spin pumping).
The first materials examined in the NLSV geometry were Al ($\alpha_\text{SH}=0.0001$ -- 0.0003) \cite{Valenzuela2006}, Pt ($\alpha_\text{SH}=0.0037$) \cite{Kimura2007} and Au ($\alpha_\text{SH}=0.11$) \cite{Seki2008}. Experiments using ISHE in combination with spin pumping reported values of $\alpha_\text{SH}=0.08$ \cite{Azevedo2011}, $\alpha_\text{SH}=0.013$ \cite{Mosendz2010PRL, Mosendz2010PRB}, $\alpha_\text{SH}=0.012$ \cite{Feng2012} for Pt and $\alpha_\text{SH}=0.0067$ \cite{Mosendz2010PRL,Mosendz2010PRB} for Au. Using the related method of spin transfer torque FMR Liu {\it{et al.}} reported  a value of $\alpha_\text{SH}=0.07$ \cite{LiuPRL2012} for Pt.
Thus depending on the experimental method used, the reported values for $\alpha_\text{SH}$ vary by more than an order of magnitude even for the same material.

The large scatter in the values of the spin Hall angles may be caused by the number of parameters used for the interpretation of the experimental data and is discussed at length in \cite{LiuCondmat2012}.
Moreover, details of the experimental configuration used to detect pure spin currents have to be considered carefully, since in particular when detecting ISHE in combination with spin pumping, parasitic signals that are often indistinguishable from ISHE may arise.
It is hence important to carefully address these issues in parallel while searching for new materials featuring large spin Hall angles \cite{Liu2012,Morota2011,NiimiPRL2012}.

In this letter the angular, material, and temperature dependence of ISHE due to spin pumping is measured.
We demonstrate that for a particular excitation geometry an unambiguous separation of signals arising from the anisotropic magneto-resistance (AMR) and ISHE is possible. Microwave frequency as well as temperature dependent measurements of ISHE, for different Py/normal metal bilayer systems are performed. We demonstrate that the amplitude of ISHE depends linearly on the microwave frequency, as predicted. Furthermore the temperature dependence of the ISHE related DC-voltage generated in Py/Pt at FMR allows the determination of the effective spin diffusion length of Pt as a function of temperature.

\begin{figure}[!h]
\begin{center}
	\includegraphics[width=0.5\textwidth]{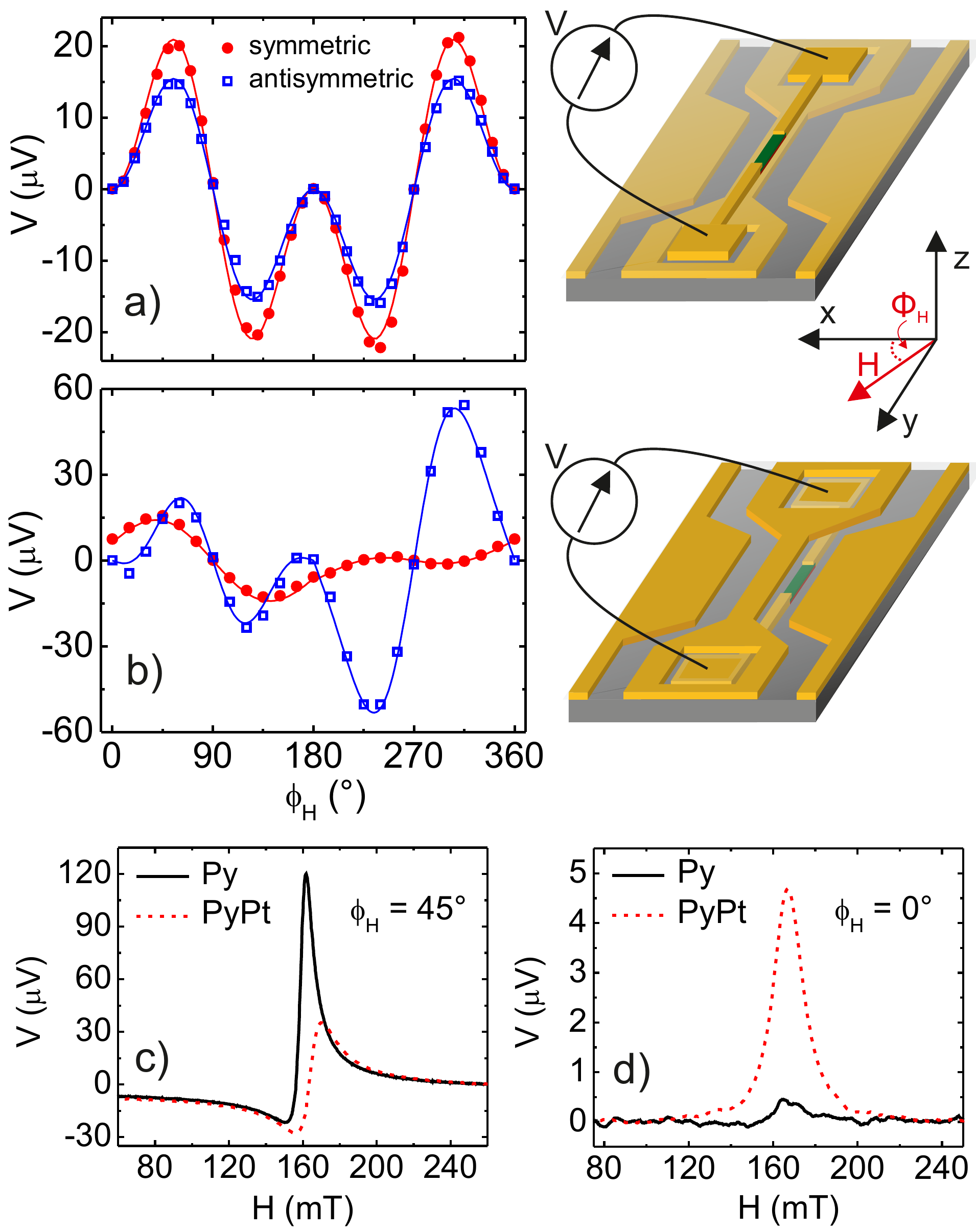}
\caption{Symmetric (red dots) and antisymmetric (blue open squares) voltage signals amplitudes at FMR (at 12 GHz) for a Py/Pt bilayer as a function of angle $\phi_\text{H}$. Corresponding sketches show the integration of a Py/Pt bilayer on top of the signal line in (a) and in the gap between signal and ground line in (b) into a CPW  leading to different excitation fields. (a) The magnetic excitation field is in-plane with respect to the Py/Pt layers. Both symmetric and antisymmetric amplitudes obey a $\sin (\phi_\text{H}) \sin (2\phi_\text{H})$ behavior as predicted in \cite{Azevedo2011}. (b) The magnetic excitation field is out-of-plane with respect to the Py/Pt layers. The amplitudes of the symmetric part follow a $\cos (\phi_\text{H}) + \sin (2\phi_\text{H})$ behavior, which reflects a contribution of both ISHE and AMR-effect to the signal. (c) Voltage around FMR measured at $\phi_\text{H}=45^\circ$, and (d) $\phi_\text{H}=0^\circ$ for a single Py layer and a PyPt bilayer. At $\phi_\text{H}=45^\circ$ both signals consist of a sum of symmetric and antisymmetric Lorentzian curves. At $\phi_\text{H}=0^\circ$ the signal for pure Py vanishes whereas the signal in PyPt is purely symmetric (pure ISHE).}
\label{AngDepIPOOP}
\end{center}
\end{figure}

The samples consist of ferromagnet/normal metal bilayers (Permalloy (Py)/Pt, Py/Au) deposited on GaAs(001) substrates. The layers are structured into wires using electron beam lithography (EBL) and sputter deposition of the single layers in ultra high vacuum. The thickness of the individual layers is \SI{12}{\nano \metre}. The bilayers are integrated into a coplanar waveguide (CPW) structure which is used to create a well defined microwave excitation field. The CPW consists of a \SI{50}{\micro \metre} wide signal line and a \SI{30}{\micro \metre} wide gap corresponding to an impedance of \SI{50}{\ohm} in the GHz frequency range. The metallic bilayer wires are \SI{5}{\micro \metre} wide and \SI{350}{\micro \metre} long  and placed either on top of the signal line (in-plane excitation) or in the gap between signal line and ground planes (out-of-plane excitation). Electrical isolation between CPW and metallic bilayer is ensured by a \SI{50}{\nano \metre} thick $\text{Al}_2\text{O}_3$ layer. The samples are placed in an external magnetic field which is rotatable in the film plane. Sketches of the samples corresponding to the different excitation field geometries are shown in Fig.~\ref{AngDepIPOOP}(a) and (b). \\
Spin current injection from a ferromagnet into an adjacent normal metal layer due to spin pumping is derived theoretically in \cite{Tserkovnyak2002PRL} and \cite{Tserkovnyak2002PRB}. It is convenient to consider the precession of the magnetization, which in the static case is saturated in the film plane, and the related injected spin current in a coordinate system ($x',y',z'$), where $x'$ always points along the direction of the dynamic magnetization, $y'$ is always in the magnetic film plane, and $z'$  points out-of-plane \cite{Celinski1997}. To describe the direction of a static magnetic field $H$ relative to the bilayer wire a second static coordinate system ($x,y,z$) is defined, as shown in Fig.~\ref{AngDepIPOOP}. Again $x$ and $y$ lie in-plane and $z$  points out-of-plane. The direction of the spin current is along $z'$ and its polarization along $x'$. In this case ISHE will generate a charge current along the $y'$-direction according to  $J_\text{C} \hat{y}' = \alpha_\text{SH} \frac{2e}{\hbar} J_\text{S}(z) \hat{z}' \times \hat{x}'$ and its magnitude is given by \cite{Azevedo2011}:\\
\begin{equation}
J_\text{C}  = \alpha_\text{SH} \frac{2e}{\hbar} J_\text{S}(0) \frac{\lambda_\text{sd}}{t_\text{nm}} \tanh \left (\frac{t_\text{nm}}{2\lambda_\text{sd}} \right ),
\end{equation}
where $\lambda_\text{sd}$ is the spin diffusion length and $t_\text{nm}$ the thickness of the normal metal layer. $J_\text{S}(0)$ is the injected spin current density at the ferromagnet/normal metal interface. The voltage probes with a separation $l$ given by the length of the Py/normal metal bilayer are placed along a fixed direction, e.g.~$y$, see Fig.~\ref{AngDepIPOOP}. The voltage due to ISHE is given by projecting $J_\text{C} \hat{y}'$ onto $\hat{y}$, integrating over $l$, and scaling by the inverse conductivity of the bilayer $1 / \sigma$. The voltage due to spin pumping and ISHE for the out-of-plane ($V_\text{ISHE}^\text{OOP}$) excitation field is then given by:
\begin{equation}
\begin{aligned}
V_\text{ISHE}^\text{OOP} &= \alpha_\text{SH}\frac{e}{\sigma}  \frac{8\pi}{M_\text{S}^2} \frac{\lambda_\text{sd}}{t_\text{nm}}l \\
&\phantom{kk} g_{\uparrow \downarrow} \omega h_\text{z}^2 \Im (\chi_\text{zz}^\text{res}) \chi_\text{yz}^\text{res} \tanh \left ( \frac{t_\text{nm}}{2\lambda_\text{sd}} \right ) \\
 & \times \left ( \frac{(\Delta H)^2}{(H - H_\text{FMR})^2 + (\Delta H)^2} \right ) \cos(\phi_\text{H}) \phantom{kkkkkkk}
\label{Vishe_OOP}
\end{aligned}
\end{equation}\\
where $H_\text{FMR}$ and $\Delta H$ are FMR field and linewidth, $\chi_\text{yz}^\text{res}$ and $\chi_\text{zz}^\text{res}$ are the in- and out-of-plane rf-susceptibilities and $\phi_\text{H}$ is the angle between $x$ and the applied magnetic field $H$, see Fig.~\ref{AngDepIPOOP}. Eq.~(\ref{Vishe_OOP}) is derived in an analogous manner as described in \cite{Azevedo2011}, but using out-of-plane excitation and the applicable entries of the magnetic susceptibility tensor. Also AMR can generate a dc-voltage signal at FMR by mixing the time dependent resistivity (AMR and precessing magnetization) with a capacitively or inductively coupled microwave current $I$ in the bilayer.
The line shape of the signal depends on the magnitude of the phase angle $\phi_\text{I}$  between precessing magnetization and rf-current in the bilayer. Due AMR the resitance of the bilayer changes with the orientation of the magnetization with respect to the current: $R_A=R_{\parallel}-R_{\perp}$. For out-of-plane excitation the generated voltage due to the AMR-effect can be described by the following formula \cite{Azevedo2011,Feng2012}:
\begin{equation}
\begin{aligned}
V_\text{AMR}^\text{OOP} & = \frac{1}{2} \frac{I R_\text{A}\chi_\text{yz}^\text{res} h_\text{z}}{M_\text{S}}\\
& \times \left ( \frac{(\Delta H)^2 \cos(\phi_\text{I})}{(H - H_\text{FMR})^2 + (\Delta H)^2  }\right. \\
	&+\left.\frac{\Delta H (H -H_\text{FMR})^{\phantom{2}}\sin(\phi_\text{I})}{(H - H_\text{FMR})^2 + (\Delta H)^2} \right ) \sin(2\phi_\text{H}) \phantom{\sin(\phi)_\text{H}},
\label{Vamr_OOP}
\end{aligned}
\end{equation}\\
First the study of a Py/Pt bilayer located on top of the signal line of a CPW (in-plane excitation), see  Fig.~\ref{AngDepIPOOP}(a), will be discussed. Around FMR of the Py layer the generated voltage signal is a Lorentzian. By fitting the aquired data to Lorentzian lineshapes the amplitudes of symmetric and antisymmetric parts of the voltage signal are determined. This is done for angles $\phi_\text{H}$ from 0 to $360^\circ$ as shown in Fig.~\ref{AngDepIPOOP}(a). The amplitudes of symmetric and antisymmetric part have the same angular dependence due to geometric reasons and can be well described by a $\sin (\phi_\text{H}) \sin (2\phi_\text{H})$ dependence \cite{Azevedo2011}. As suggested in \cite{Azevedo2011} the phase angle $\phi_\text{I}$ is not necessarily equal to $90^\circ$ at FMR. Therefore, the amplitude of the symmetric part of the voltage arises in general from a sum of signals originating from ISHE and AMR. It is thus not possible to uniquely separate signals arising from ISHE from signals caused by AMR using in-plane magnetic field excitation. However, from Eq.~(\ref{Vishe_OOP}) it becomes clear that by placing a bilayer in the gap between signal and ground line of a CPW, see the sketch in Fig.~\ref{AngDepIPOOP}(b), and measuring at $\phi_\text{H}=0^\circ$ only ISHE contributes to the voltage signal. The angular dependence of the amplitudes of symmetric and antisymmetric part of the voltage in Fig.~\ref{AngDepIPOOP}(b) shows, that the symmetric part is generally a mixture of ISHE and AMR-effect related signals. The amplitude of the antisymmetric part vanishes when $\phi_\text{H}$ is an integer multiple of $180^\circ$. At these angles the voltage signal is purely symmetric and can only be caused by ISHE. The asymmetry of the data presented in  Fig.~\ref{AngDepIPOOP}(b) with respect to $\phi_\text{H}=180^\circ$  is most likely caused by the phase angle $\phi_\text{I}$. The data shown in Fig.~\ref{AngDepIPOOP}(b) are consistent with $\phi_\text{I}=40^\circ$ for $\phi_\text{H}=0^\circ - 180^\circ$. The angular plot of Fig.\ref{AngDepIPOOP}(b) suggests, that the phase $\phi_\text{I}$ between capacitively coupled current and precessing magnetization is not constant, but shifts across $\phi_\text{H}=180^\circ$ to a corresponding value of $\phi_\text{I}=80^\circ$. This is also reflected in the fact that the amplitude of the symmetric voltage part for $\phi_\text{H}=0^\circ - 180^\circ$ is larger than for $\phi_\text{H}=180^\circ - 360^\circ$. The phase difference for $\phi_\text{I}$ corresponds to an excitation field angle of $20^\circ$ relative to the normal of the bilayer \cite{Bai2013}. In Fig.~\ref{AngDepIPOOP}(c) and (d) we  compare of voltage signals at FMR for a single Py layer (where one only has AMR and no ISHE) and a Py/Pt bilayer. At $\phi_\text{H}=45^\circ$ both samples show a similar lineshape with symmetric and antisymmetric contribution. The large symmetric signal amplitude for pure Py demonstrates that the phase angle $\phi_\text{I}$ is not $90^\circ$ in our devices. However at $\phi_\text{H}=0^\circ$ the voltage signal in Py vanishes whereas the signal in PyPt is purely symmetric around the FMR-field. Clearly the AMR contribution vanishes at $\phi_\text{H}=0^\circ$.

\begin{figure}[!h]
\begin{center}
	\includegraphics[width=0.5\textwidth]{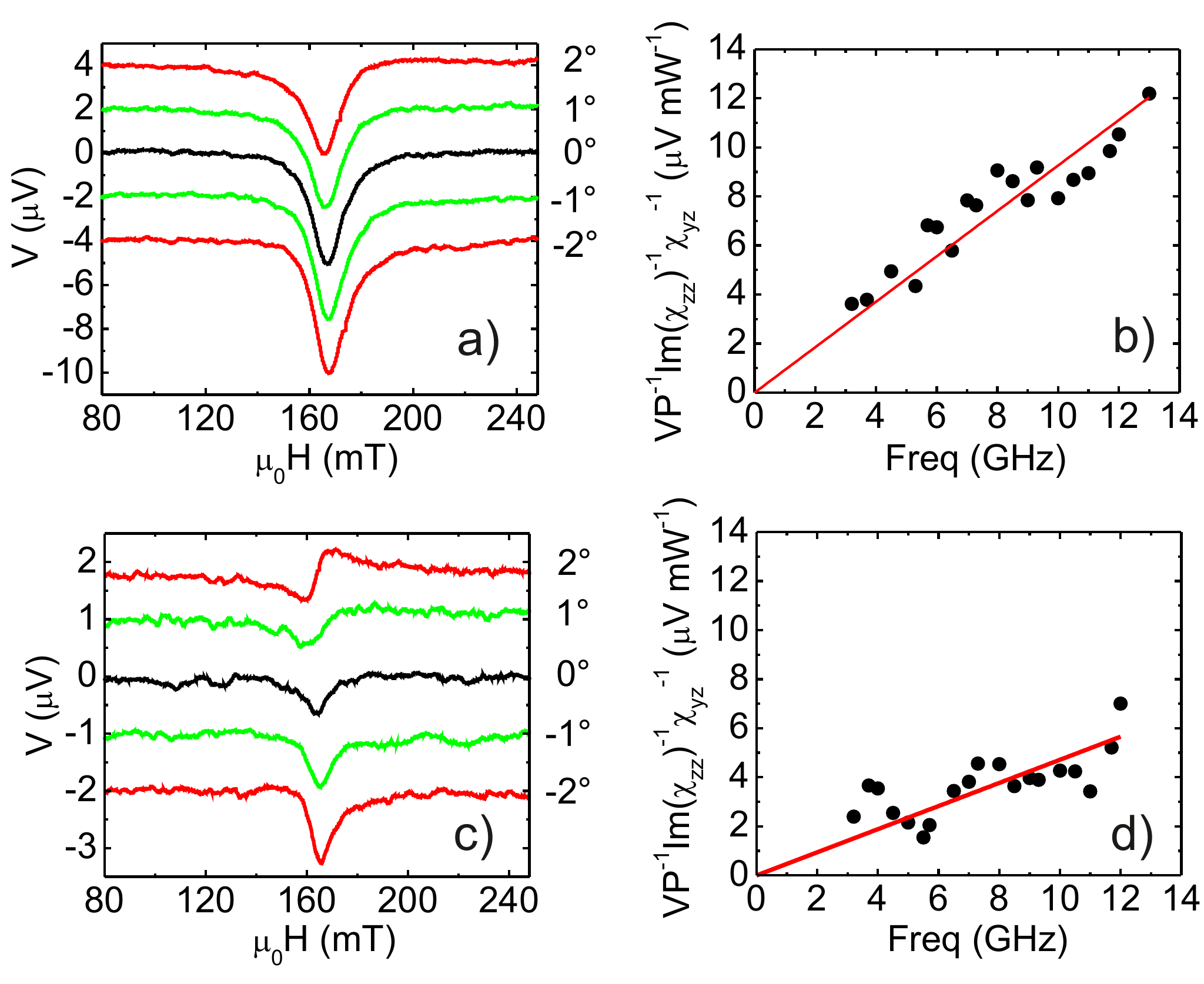}
\caption{Voltage signals at FMR (at 12 GHz) for Py/Pt and Py/Au at small angles around $\phi_\text{H}=0^\circ$, panel a) and c) respectively. Normalizing with respect to frequency dependent parameters (microwave power, susceptibility and ellipticity) the amplitude of the signal exhibits a linear frequency dependence at $\phi_\text{H}=0^\circ$ for both Py/Pt and Py/Au, panel b) and d) respectively.}
\label{FreqDep}
\end{center}
\end{figure}

Next we focus on the pure ISHE signal measured at $\phi_\text{H}=0^\circ$. Eq.~(\ref{Vishe_OOP}) suggests that a signal due to ISHE should scale linearly with the microwave frequency. This was verified  by performing frequency dependent measurements for Py/Pt and Py/Au bilayers. First  $\phi_\text{H}=0^\circ$ is determined precisely by stepping the magnetic field angle and recording the signals as shown in Figs.~\ref{FreqDep} (a) and (c) for Py/Pt and Py/Au, respectively. In Figs.~\ref{FreqDep} (b) and (d) the results for the frequency dependence of the ISHE for Py/Pt and Py/Au are shown. These data are normalized with respect to the frequency dependent susceptibility and microwave transmission of the sample. First, the data have to be divided by the factor $\Im (\chi_\text{zz}^\text{res}) \chi_\text{yz}^\text{res}$, which contains the susceptibilities and takes the ellipticity of the precession into account. The normalization of the data with respect to the microwave transmission of the sample is done using the average of input and output power as the normalization factor.
The spin Hall angles for Pt and Au can be calculated according to the following formula which is the result of solving Eq.(\ref{Vishe_OOP}) for $\alpha_\text{SH}$ at FMR.
\begin{equation}
\alpha_\text{SH} = \frac{V_\text{ISHE}^\text{OOP} \sigma_\text{nm} M_\text{S}^2 t_\text{nm}}{8 \pi e \lambda_\text{sd} g_{\uparrow \downarrow} \omega h_\text{z}^2 \Im(\chi_\text{zz}^\text{res})\chi_\text{yz}^\text{res} l \tanh \left ( \frac{t_\text{nm}}{2\lambda_\text{sd}} \right )},
\label{spinHallAngle}
\end{equation}
where $ g_{\uparrow \downarrow}$ is the spin mixing conductance. This parameter describes how much spin current is generated by spin pumping and enters the normal metal \cite{Tserkovnyak2002PRB}. For Pt a $g_{\uparrow \downarrow}$ value of \SI{2.5e15}{\per \square \centi \metre} is determined by measuring the line broadening due to the presence of a Pt cover layer which is in agreement with presented data in \cite{Tserkovnyak2002PRB} and \cite{Mizukami2001}. For Au a value of \SI{1.2e15}{\per \square \centi \metre} is used \cite{Mosendz2009PRB}. The out-of-plane excitation field was determined to be \SI{0.24}{\milli \tesla} at 100~mW using electromagnetic simulations of the CPW structure. Furthermore the spin diffusion length enters Eq.~(\ref{spinHallAngle}). For this parameter recent spin pumping experiments suggest a value of \SI{1.4}{\nano \metre} \cite{LiuPRL2012,LiuCondmat2012} for Pt. For Au a spin diffusion length of \SI{35}{\nano \metre} was measured in \cite{Mosendz2009PRB}. Taking into account all data shown in Fig.~\ref{FreqDep}  and $\lambda_\text{sd}=\SI{1.4}{nm}$ a spin Hall angle of $\alpha_\text{SH}$ of $0.09\pm 0.02$ is obtained for Pt. Similarly using $\lambda_\text{sd}=\SI{35}{\nano \metre}$ a value of $0.008\pm 0.001$ is deduced for Au.
\begin{figure}[!h]
\begin{center}
	\includegraphics[width=0.5\textwidth]{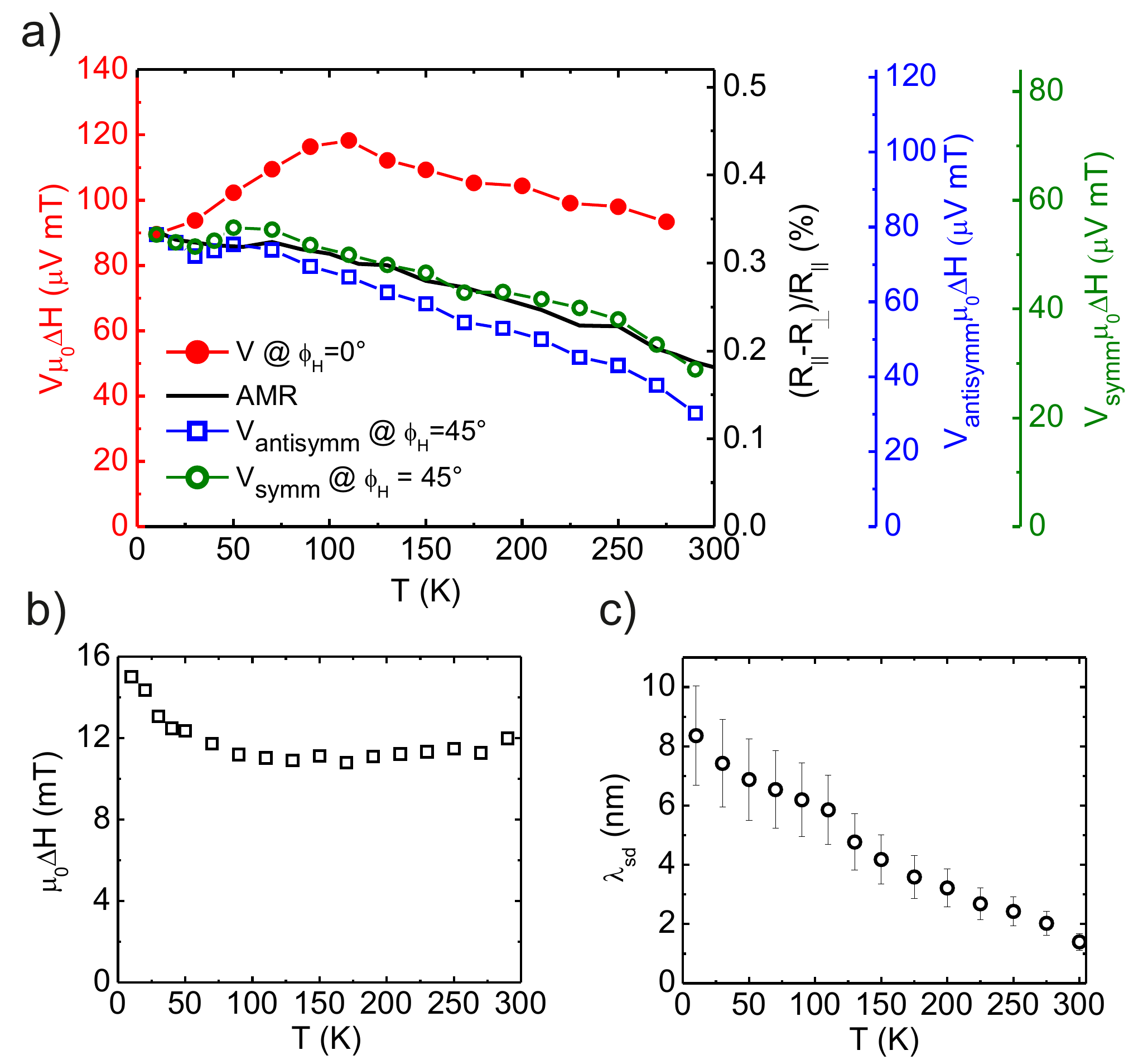}
\caption{(a): Signal amplitudes measured at FMR (at 18 GHz) in Py/Pt multiplied with the respective temperature dependent FMR-linewidths for $\phi_\text{H}=0^\circ$ and $\phi_\text{H}=45^\circ$ are plotted. The red curve (filled dots) is the signal at  $\phi_\text{H}=0^\circ$, the green (empty dots) and the blue curve (empty squares) represent the symmetric and the antisymmetric part of the voltage at $\phi_\text{H}=45^\circ$. The black curve (line) is the AMR measured statically in the same sample. (b) Temperature dependence of the FMR-linewidths. (c) Temperature dependence of the spin diffusion length calculated from the signal measured at $\phi_\text{H}=0^\circ$ assuming a constant $\alpha_\text{SH}=0.09\pm0.02$ as calculated for room temperature experiments.
}
\label{TDep}
\end{center}
\end{figure}

In order to demonstrate the different physical origin of the signals that are obtained at $\phi_\text{H}=0^\circ$ (ISHE) and $\phi_\text{H}=45^\circ$ (AMR+ISHE) the temperature dependence of the voltage signals was measured for a  Py/Pt bilayer. The results are displayed in Fig.~\ref{TDep}. In Fig.~\ref{TDep}(a) the voltage signals at  $\phi_\text{H}=0^\circ$  and $\phi_\text{H}={45^\circ}$ are compared to the static AMR measured in the same Py/Pt bilayer sample. The static AMR is determined by resistance measurement on the bilayer for parallel ($\parallel$) and perpendicular ($\perp$) configurations of the external magnetic field  with respect to the current path. Consequently $R_\perp$ is subtracted from $R_\parallel$ and the result is divided by $R_\parallel$.
 In order to interpret the physical origin of the different signals in Fig.~\ref{TDep}(a) the data sets are scaled to reach the same value at T=\SI{10}{\kelvin} and normalized by the temperature dependent FMR-linewidths (shown in Fig.~\ref{TDep}(b)). While the normalized signals measured at $\phi_\text{H}=45^\circ$ follow the temperature dependence of the static AMR, the signal at $\phi_\text{H}=0^\circ$ has a different temperature dependence. The AMR-effect generates a voltage which is proportional to the inverse resistance of the Py-layer (implicitly contained in the capacitively coupled microwave current \cite{Mosendz2010PRB}) and is inversely proportional to the FMR-linewidth. The different temperature dependencies shown in Fig.~\ref{TDep}(a) strongly suggest that the voltage measured for $\phi_\text{H}=45^\circ$ is mainly due to the AMR-effect while the signal at $\phi_\text{H}=0^\circ$ only is purely caused by ISHE.  According to theory the voltage due to ISHE  is proportional to $1/(\Delta H^2 \sigma)$. Using the fact that the spin Hall angle of Pt is $0.09\pm0.02$ at room temperature where we use $\lambda_\text{sd}=\SI{1.4}{\nano \metre}$ and the assumption, that the spin Hall angle is constant with temperature, the temperature dependence of the spin diffusion length for Pt can be calculated using Eq.~(\ref{spinHallAngle}). The result is shown in Fig.~\ref{TDep}(c).\\

In conclusion, we show that it is not possible to reliably measure ISHE separately from the parasitic AMR for in-plane excitation of the magnetization vector in a Py/normal metal bilayer using a coplanar waveguide experiment. Only out-of-plane excitation and only at certain angles of the polarization of the injected spin current with respect to the voltage probes an unambiguous study of ISHE is possible. Using this geometry spin Hall angles can be quantified reliably using parameters inferred from spin pumping measurements. Temperature dependent measurements underpin the anisotropic magnetoresistive nature of parasitic voltage signals. Moreover, it strongly suggests, that the origin of the signals changes from a mixture of signals arising from ISHE and AMR-effect to a signal originating from ISHE only at appropriate angles. Moreover, the temperature dependent spin diffusion length of Pt has been determined using the inverse spin Hall effect.

\FloatBarrier

\bibliography{literature1}
\bibliographystyle{unsrt}
\end{document}